\documentclass[10pt]{article}
\usepackage{latexsym,graphicx}
\usepackage{latexsym,graphicx}
\newcommand{\be}{\begin{equation}}
\newcommand{\ee}{\end{equation}}
\usepackage[left=2cm,top=2.5cm,right=2.5cm,bottom=1.5cm]{geometry}

\catcode `\@=11 \catcode `\@=12
\begin{document}
\begin{center}
\large{\bf {REFRACTION-BASED ALTERNATIVE EXPLANATION FOR: BENDING OF LIGHT NEAR A STAR, GRAVITATIONAL RED/BLUE SHIFT AND BLACK-HOLE}} \\
\vspace{8mm}
\normalsize{R. C. GUPTA $^1$, ANIRUDH PRADHAN $^2$ and SUSHANT GUPTA $^3$} \\
\vspace{5mm}
\normalsize{$^{1}$ Institute of Engineering and Technology (IET), AKTU(UPTU), Lucknow-226 021, India\\
E-mail: rcg\_iet@hotmail.com}\\
\vspace{5mm} \normalsize{$^{2}$ Department of Mathematics, Institute of Applied Sciences and Humanities, GLA University, \\
Mathura-281 406, India\\
E-mail: pradhan@iucaa.ernet.in}\\
\vspace{5mm} \normalsize{$^{3}$ Department of Physics, Lucknow University, Lucknow 226 007, India\\
E-mail: sushant1586@gmail.com}\\
\end{center}
\vspace{9mm}
%\date{}
\begin{abstract}
In this research-paper, many of the general-relativity-tests such as bending of light near a star and gravitational
red/blue shift are explained {\it without} general-relativity \& even {\it without} Newtonian-approach. The authors
first raise questions on the validity of both, the Newtonian and the relativistic approach; and then propose a novel
alternative-explanation. The new alternative explanation is based on refraction-phenomenon of optics. Estimation of
results with new approach are in agreement with known values. Though physics is different, but it is argued that
general-relativity based gravitational-bending and refraction based bending have more in common than is generally
realized. Also discussed are black-hole and gravitational-lensing in the new perspective of refraction. The new
refraction-based theory makes a few new predictions and also suggests a few tests.
\end{abstract}
%\newpage

%%%%%%%%%%%%%%%%%%%%%%%%%%%%%%%%%%%%%%%%%%%%%%%%%%%%%%%%%%%%%%%%%%%%%%%%%
%%%%%%%%%%%%%%%%%%%%%%%%%%%%%%%%% SECTION 1 %%%%%%%%%%%%%%%%%%%%%%% %%%%%
\section{INTRODUCTION}
As observed on Earth, light from a distant star/planet (such as Venus) bends when it passes near another star (such as Sun).
Einstein's theory predicts $(\frac{4GM}{c^{2}R})$ double the bending as predicted $(\frac{2GM}{c^{2}R})$ by conventional
Newtonian mechanics. Experimental confirmation was the triumph of Einstein's general theory of relativity [1-3]. In the
present paper, the authors, however, first raise questions on the validity of both - the Newtonian explanation and Einstein's
explanation in the two subsequent paragraphs; and then propose an alternative explanation based on refraction-phenomenon of
optics. The alternative theory in this paper also explains black-hole and gravitational red/blue-shift.\\

Conventional Newtonian explanation for bending of light is based on photon's gravitational attraction towards the star (Sun).
In fact, photon has no material mass (rest-mass zero) and has mass $(\frac{h\nu}{c^{2}})$ only due to its energy $(h\nu)$. It
is understood that gravitation is only between material bodies; the authors raise questions on the validity of Newtonian
gravitational-attraction on photon. Though photon has energy \& momentum \cite{ref4}; it does not seem to have inertial \&
gravitational mass, else it would have been possible to accelerate or decelerate it, and all the photons in a Light-beam would
focus by themself but such an auto-focusing effect has never been observed \cite{ref5}. Moreover, if photon is considered as wave,
it is not clear as how ( \& if ) gravity can influence it ?. The authors thus conclude that gravitation (of Sun) does not influence
photon (coming from Venus), therefore can not cause bending of light through Newtonian-mechanics. Also thus gravitation should
not be responsible for the so-called gravitational-red/blue-shift.\\

Einstein's general-relativity explanation is based on geodesic or curvature of space-time near a massive body. Although
general-relativity has passed several tests, but it has a serious weak point. According to Einstein, gravity is
`not a real force', but an artifact of curvature. But our everyday physical experience is contrary to that  and tells
that gravity is a real-force. `Gravity is a real-force and could have electrostatic-origin' is rather indirectly confirmed
through the famous Millikan's oil-drop experiment \cite{ref6} wherein the gravitational-force is balanced with the Coulomb's
repulsion on the electro-statically charged tiny oil-drops. Some scientists around the globe such as Gupta \cite{ref7},
Ron Kita \cite{ref8} and John Newell \cite{ref9} and many others [6, 10-15] also believe that Gravity is a real-force and
has electro-static origin; a few of them [6, 8, 10, 15] have even shown some experimental indications of electro-gravity. Most
of the learned scientists, however, are tuned to what they have been taught, i.e., `general-relativity' !   It seems that
Prof. Everett has rightly quoted \cite{ref16} a letter from Thorne \& Will that : ``Physicists attitude about gravity have
been conditioned by general-relativity. . . . However, we see no reasons why Nature should conform to the present convenience
of physics. If  She (the Nature) has chosen to go a {\it different route} from general-relativity this will shake the foundations
of physics''. The {\it different route} from general-relativity could be the one suggested by Gupta \cite{ref7}: `Gravity as the
second-order manifestation of electrostatic-force {\it via}  special-relativity', in the same-way as magnetism is known to be
considered as the first-order manifestation of electricity {\it via}  special-relativity. The present authors wish to mention that
if Ptolemy geo-centric model can ultimately change after thirteen centuries, Newton's model can be modified after three centuries,
Einstein's general-relativity theory too can also be altered in one century. Scientific \& political history teaches us many
lessons; one such lesson is that the `inertia' not only lies  with `matter'  but much-more with `gray-matter (brain)' implying
that our learned-people are more reluctant to accept the new models. This unfortunately hinders the scientific-developments.
Thus, for the sake of Truth to prevail, all new avenues for possible Truth must be kept open. Einstein himself has also said
\cite{ref17} that ``to raise new questions, new possibilities, to regard old problem from a new angle, requires creative
imaginations and makes real advance in science''. \\

The authors propose alternative explanation based on refraction-phenomenon of optics, for bending of light near star.
When Venus-light from space (say, vacuum or near-vacuum) enters into the star's surroundings/atmosphere (comparatively
denser medium) the light-ray bends towards the star Sun) due to common well known phenomenon of refraction of light.
The refraction-based theory is also able to explain gravitational red/blue shift. Also, black-hole, gravitational-lensing
and space-time too are considered in the new perspective. In view of the uncertainty \& unavailability of information/data
regarding refractive-index of atmospheric-medium and its variations; a rather semi-empirical approach, for the alternative
explanation for bending of light near a star and gravitational red/blue shift etc., is appropriate and is described
in the paper as follows.

%%%%%%%%%%%%%%%%%%%%%%%%%%%%%%%%%%%%%%%%%%%%%%%%%%%%%%%%%%%%%%%%%%%%%%%%%
%%%%%%%%%%%%%%%%%%%%%%%%%%%%%%%%% SECTION 2 %%%%%%%%%%%%%%%%%%%%%%% %%%%%
\section{A BRIEF MENTION OF EARLIER REFRACTION-BASED THEORIES FOR BENDING OF LIGHT NEAR A STAR}
Only some scanty \& scattered informations are available in literature \& on internet, such as a few recent ones
quoted [18-26] here, on refraction-based theories (or alternative theories) for bending of light near a star. However,
unfortunately, the alternative attempts seem to have been met with less-appreciation and more-ridicules. Many people
were even hesitant to propose such alternative theories against well-established general-relativity-based
gravitational-bending, in fear of being criticized \& ridiculed. Nevertheless, some [22-26] have dared to go
a step further and even questioned the established notions of physics, and that some [24-26] have suggested that
most of the `general-relativity test' could be explained even {\it without} `general-relativity'. \\

Many alternative theories of gravitation have been proposed in literature, can also be found on net, not discussed
herein, however, a brief mention of it can be found in the reference \cite{ref7} and the references therein, but
it is to mention that mostly these are `metric-based' \& variant of the general-relativity. A novel theory of gravitation,
as an alternative to Einstein's General-theory-of-gravitation, proposed by Gupta \cite{ref7} interestingly relies on
Einstein's Special-theory-of-relativity; Einstein's (general-relativity theory) view and the author's (alternative gravity theory)
view on `curvature' of space-time are discussed in sections-7 \& 8 of this paper, both the views appear to be different but
the essence \& effect are more-or-less the same. Truly, some [25,26] agree that: it can be argued that bending due to
light-refraction and general-relativity based gravitational-bending have more in common (as explained in the section-8)
than is generally realized.

%%%%%%%%%%%%%%%%%%%%%%%%%%%%%%%%%%%%%%%%%%%%%%%%%%%%%%%%%%%%%%%%%%%%%%%%%%%%%%%%%%%%%%%%%%%%%%%%%%%%%%%%%%%%
%%%%%%%%%%%%%%%%%%%%%%%%%%%%%%%  SECTION 3  %%%%%%%%%%%%%%%%%%%%%%%%%%%%%%%%%%%%%%%%%%%%%%%%%%%%%%%%%%%%%%%%
\section{BENDING OF LIGHT NEAR A STAR: THE ALTERNATIVE EXPLANATION}

%%%%%%%%%%%%%%%%%%%%%%%%%%%%%%%%%%%%%%%%%%%%%%%%%%%%%%%%%%%%%%%%%%%%%
%%%%%%%%%%%%%%%%%%%%%%%%%%%%%%%  SUBSECTION 3.1  %%%%%%%%%%%%%%%%%%%%
\subsection{THE PRINCIPLE: DEVIATION DUE TO REFRACTION}
Refraction of light rays is a well known optics-phenomenon \cite{ref27}. This provides an alternative explanation of bending of
light near a star. When light ray, from space (near vacuum), enters the star's atmosphere (medium); the light ray bends near the
star due to refraction. To illustrate the bending due to refraction, consider a spherical water-droplet as shown in Figure(1.a).
When light ray enters from lighter medium (air) to denser medium (water), the droplet works as prism and thus the light-ray bends
due to refraction. Similarly, when light ray enters from space-vacuum (lighter medium) to star's atmosphere (denser medium) it
bends due to refraction as shown in Figure(1.b). The atmosphere extends to great heights, it becomes rarer and rarer, however;
a reasonable equivalent height is shown in the figure. \\

The amount of bending (maximum deviation) can be estimated semi-empirically $( \delta = 2(\mu -1)$ as shown in section $3.3$)
as follows. Consider the limiting case when the light ray enters the atmosphere touching at point $A$ and leaves touching at
point $C$. The incident ray touches at $ i = 90^{0} $ \& refraction angle is $r$ at point $A$ and vice-versa at point $C$
as shown in Figure- $2$. The angle $r$ is thus critical angle $(\mu = \frac {1}{sin r})$, and for maximum deviation the line
$AC$ touches the star-core at point $B$. From the star-geometry of Figure-$2$, $ Cosec (r) = \frac {R^{/}}{R}$ where $R^{/}$
\& $R$ are atmospheric-radius and core-radius of the star. Thus $ \mu = \frac {R^{/}}{R} = \frac{(R + h)}{R} = 1 + \frac{h}{R}$
where $h$ is the equivalent-height of atmosphere above the star-core (estimated in the following section $3.2$). For max deviation
situation (Figure-$2$) thus,
\begin{equation}
\label{eq1}
\mu = Cosec (r) = \frac{R^{/}}{R} = 1 + \frac{h}{R}
\end{equation}

%%%%%%%%%%%%%%%%%%%%%%%%%%%Figure1%%%%%%%%%%%%%%%%%%%%%%%%%
%%%%%%%%%%%%%%%%%%%%%%%%%%%%%%%%%%%%%%%%%%%%%%%%%%%%%%%%%%%
\begin{figure}[htbp]
\centering
\includegraphics[width=8cm,height=10cm,angle=0]{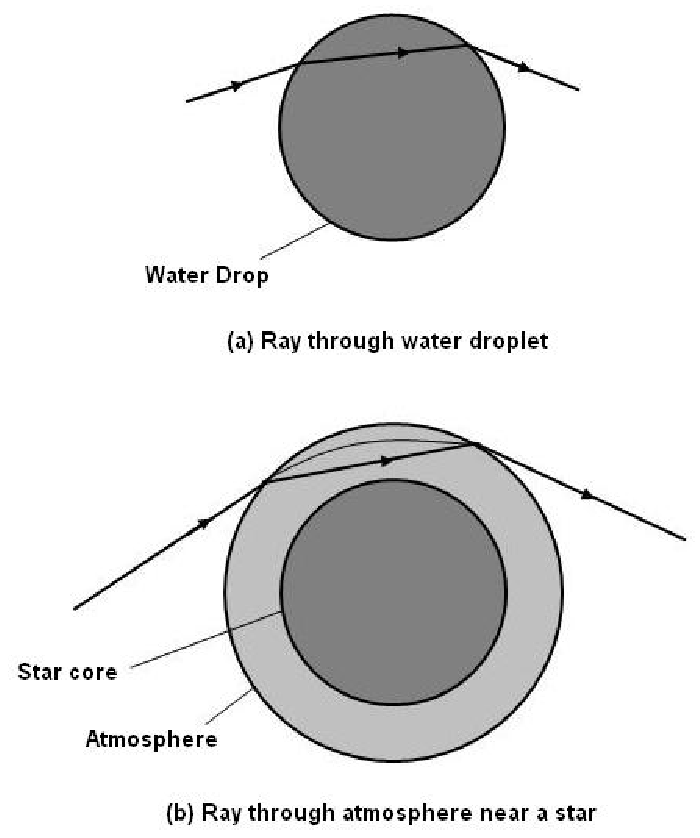}
\textbf{\caption{\textit{\textsl{Refraction of light}}}}
\end{figure}
%%%%%%%%%%%%%%%%%%%%%%%%%%%%%%%%%%%%%%%%%%%%%%%%%%%%%%%%%%%
%%%%%%%%%%%%%%%%%%%%%%%%%%%%%%%%%%%%%%%%%%%%%%%%%%%%%%%%%%%

%%%%%%%%%%%%%%%%%%%%%%%%Figure2%%%%%%%%%%%%%%%%%%%%%%%%%%%%
%%%%%%%%%%%%%%%%%%%%%%%%%%%%%%%%%%%%%%%%%%%%%%%%%%%%%%%%%%%
\begin{figure}[htbp]
\centering
\includegraphics[width=10cm,height=10cm,angle=0]{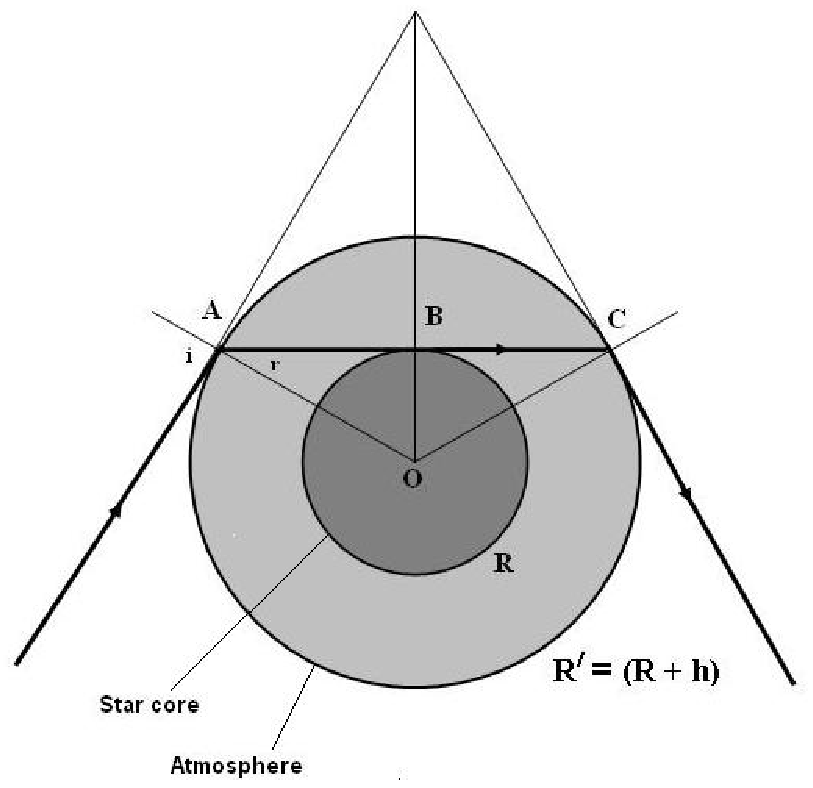}
\textbf{\caption{\textit{\textsl{Bending of light near a star}}}}
\end{figure}
%%%%%%%%%%%%%%%%%%%%%%%%%%%%%%%%%%%%%%%%%%%%%%%%%%%%%%%%%%%
%%%%%%%%%%%%%%%%%%%%%%%%%%%%%%%%%%%%%%%%%%%%%%%%%%%%%%%%%%%%%%%%%%%%%
%%%%%%%%%%%%%%%%%%%%%%%%%%%%%%%  SUBSECTION 3.2  %%%%%%%%%%%%%%%%%%%%
\subsection{SEMI-EMPIRICAL ESTIMATIONS OF EQUIVALENT HEIGHT OF THE ATMOSPHERE
$h$ AND ITS AVERAGE REFRACTIVE INDEX $\mu$}
Star-core is enveloped by dense \& diluted gaseous surroundings (or atmosphere) with varying density \& refractive-index.
It is thus difficult to estimate the equivalent height $h$ of star's atmosphere, within which properties are assumed to
be uniform. Factors such as gravitation, temperature, pressure, density, radiation-pressure etc., can influence it. But
the main factors for star (Sun) seems to be gravitation and radiation-pressure. The dimensionless ratio $\frac{h}{R}$
could probably depend (proportional) on another dimensionless quantity $\frac{GM}{c^{2}R}$ (mentioned earlier for bending
of light near the star). This in a way takes into account gravitation (gravitational potential energy - $\frac{GM}{R}$) and
radiation (velocity of light $c$). Thus taking (assuming) proportionality-constant or fuzz-factor as $k$, the star's equivalent
atmospheric-height $h$ is taken semi-empirically as:
\begin{equation}
\label{eq2}
\frac{h}{R} = \frac{kGM}{c^{2}R}
\end{equation}
In optics, the refractive index of a medium $ \mu = \frac{\sin
i}{\sin r}$ is also known as ratio of velocity of light in vacuum to
the medium, i.e., $ \mu = \frac {c_{o}}{c_{m}} $. Also since
velocity of light (electro-magnetic wave) $c = \frac{1}{(\epsilon
u)^{\frac{1}{2}}}$ where $\epsilon$ and u are electric-permittivity
and magnetic-permeability of the medium; $ \mu = (\epsilon_{r}
u_{r})^{\frac{1}{2}} \approx (\epsilon_{r})^{\frac{1}{2}}$ as
relative-permeability $u_{r} \approx 1$, where $\epsilon_{r}$ is
average relative-permittivity (dielectric constant of the medium)
which itself is given as $\epsilon_{r} = 1 + \chi$ where $\chi$ is
the average electric-susceptibility of the medium \cite{ref28}. Thus
$(\epsilon_{r})^{\frac{1}{2}} = (1 + \chi)^\frac{1}{2} \approx 1 +
\frac{\chi}{2}$. Though $\chi$, $\epsilon_{r}$ \& $\mu$ vary within
the atmosphere with maximum at star-core to minimum at outer-layer
of atmosphere; but considering the average values of $\epsilon_{r}$
\& $\mu$, the average equivalent value of $\mu$ (Eq. \ref{eq3}) and
$\chi$ (Eq. \ref{eq4}, using Eqs. \ref{eq1}, \ref{eq2} \& \ref{eq3})
are given as,
\begin{equation}
\label{eq3}
\mu = 1 + \frac{\chi}{2}
\end{equation}
\begin{equation}
\label{eq4}
\chi = \frac{2kGM}{c^{2}R}
\end{equation}
%%%%%%%%%%%%%%%%%%%%%%%%%%%%%%%%%%%%%%%%%%%%%%%%%%%%%%%%%%%%%%%%%%%%%
%%%%%%%%%%%%%%%%%%%%%%%%%%%%%%%  SUBSECTION 3.3  %%%%%%%%%%%%%%%%%%%%
\subsection{ESTIMATION OF BENDING (DEVIATION) OF LIGHT NEAR A STAR DUE TO REFRACTION-PHENOMENON}

The angular deviation at entry point $A$ (Figure(2)) is $(i - r)$, and
similar deviation of the ray occurs at exit point C. So, the total
deviation (bending) $\delta = 2(i - r)$. From optics consideration
and using simplification \& approximation, and also noting that
deviation is more for higher $\mu$ \& that there is no-deviation for
$\mu = 1$; it can be shown that deviation $(i - r)$ $\approx (\mu -
1)$. Hence the expressions for total deviation $\delta$ are given as
in Eq. \ref{eq5}, as in Eq. \ref{eq6} (using Eqs. \ref{eq5} \&
\ref{eq3}) and as in Eq. \ref{eq7} (using Eqs. \ref{eq6} \&
\ref{eq4}):
\begin{equation}
\label{eq5}
\delta = 2(\mu - 1)
\end{equation}
\begin{equation}
\label{eq6}
= \chi
\end{equation}
\begin{equation}
\label{eq7}
= \frac{2kGM}{c^{2}R}
\end{equation}
The total deviation (bending of light) $\delta =
\frac{2kGM}{c^{2}R}$ given by Eq. \ref{eq7} is same (for fuzz factor
$k=2$) as that predicted by the celebrated general-relativity and
found experimentally correct. The approach (physics) of the present
explanation, however, is altogether different and is much simpler.
The new approach is based on the commonly well-known phenomenon of
refraction of light; there is, however, a fuzz-factor k to account
for uncertainty such as in estimation of star's atmospheric height
\& its refractive index. The authors aim to emphasize that though
refraction-phenomenon approach and general-relativity approach are
in agreement as far as result is concerned but the physics of both
the approaches are quite different.

%%%%%%%%%%%%%%%%%%%%%%%%%%%%%%%%%%%%%%%%%%%%%%%%%%%%%%%%%%%%%%%%%%%%%
%%%%%%%%%%%%%%%%%%%%%%%%%%%%%%%  SUBSECTION 3.4  %%%%%%%%%%%%%%%%%%%%
\subsection{GRAVITATIONAL-LENSING (IN NEW LIGHT AS REFRACTION-BENDING)}

In perspective of refraction phenomenon discussed for bending of
light, the so called gravitational-lensing \cite{ref29} is in fact
`real' refraction-lensing of light due to refraction through
atmospheric-layer of star or galaxy (note-both star \& galaxy are
surrounded with cloud of gases/materials, both can cause
refraction-bending of light and thus lensing). In fact the word
`lensing' here literally means real lensing (bending of light due to
refraction). But through optical-lens deviation occurs with some
dispersion too, causing chromatic aberration. It is expected that
here too, if the lensing is due to refraction (as said in the
present paper), a little dispersion (chromatic aberration) can also
occur which may possibly be found experimentally. The sky as if will
look more colorful, and it is the color which will differentiate
between the object \& its image.

%%%%%%%%%%%%%%%%%%%%%%%%%%%%%%%%%%%%%%%%%%%%%%%%%%%%%%%%%%%%%%%%%%%%%%%%%%
%%%%%%%%%%%%%%%%%%%%%%%%%%% SECTION 4 %%%%%%%%%%%%%%%%%%%%%%%%%%%%%%%%%%%%
\section{A NEW LOOK ON BLACK-HOLE: LIGHT-TRAPPING INSIDE, DUE TO TOTAL-INTERNAL-REFRACTION}

What happens if the light ray after entering the atmosphere (at $r$ =
critical angle) suffers total-internal-reflection (due to slight
change in refractive index or angle) at point C (Figures-$2$ \& $3$).
The ray will thus continue to travel inside the atmosphere along a
closed regular polygon as shown in Figure-$3$. For max possible
deviation \& nearness to star, the ray enters at critical-angle and
touches the star-core at point $B$ (Figure-$3$). If the ray $AC$ passes
above $B$, it will not suffer total-internal-reflection and thus will
come out of atmosphere at the first-pass at point $C$; whereas if ray
sight $AC$ is below $B$ the ray will be trapped (obstructed, thus
absorbed) by the star-core.\\

%%%%%%%%%%%%%%%%%%%%%%%%%%%Figure3%%%%%%%%%%%%%%%%%%%%%%%%%%%%%
%%%%%%%%%%%%%%%%%%%%%%%%%%%%%%%%%%%%%%%%%%%%%%%%%%%%%%%%%%%%%%%
\begin{figure}[htbp]
\centering
\includegraphics[width=8cm,height=18cm,angle=0]{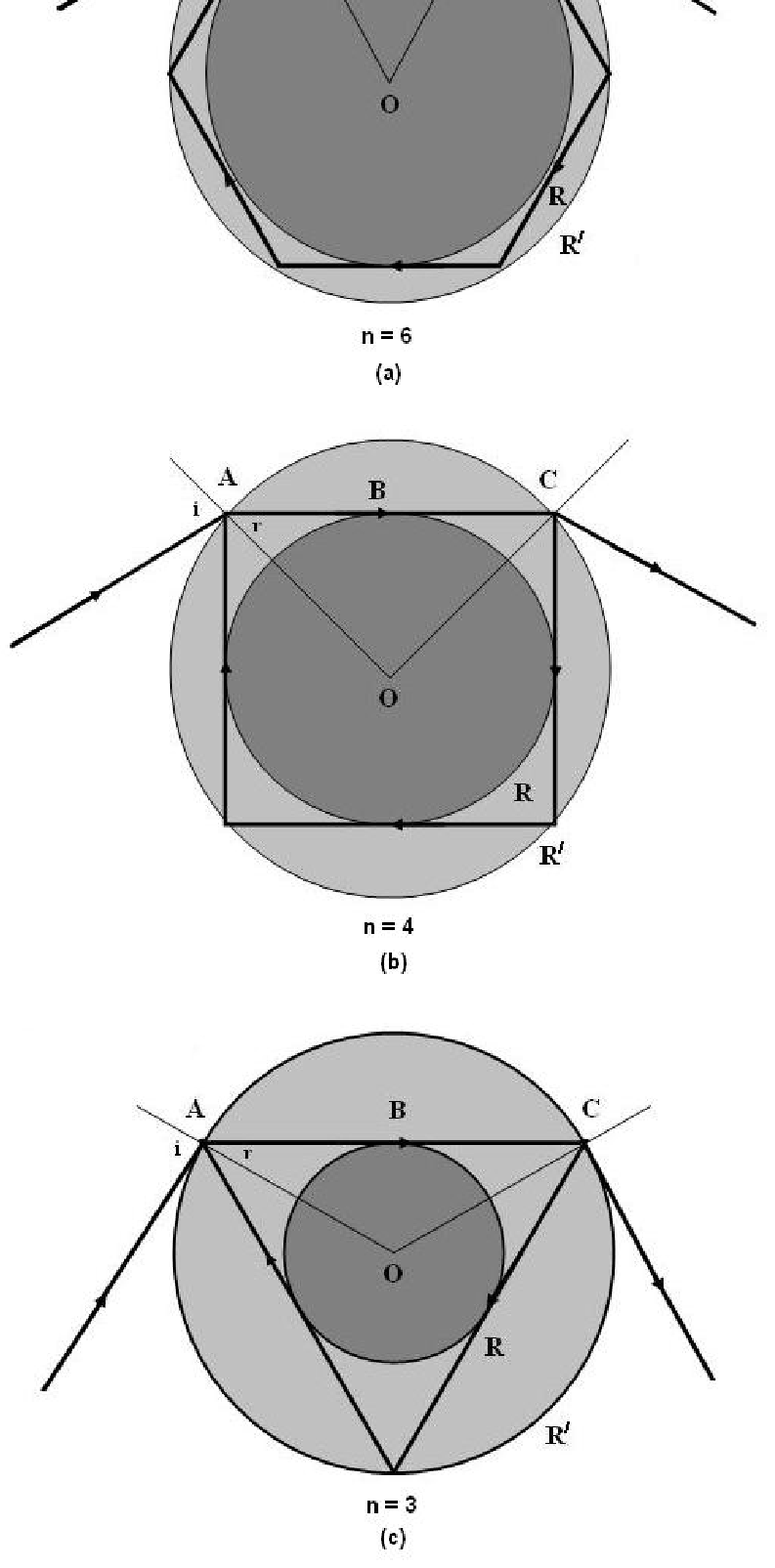}
\textbf{\caption{\textit{\textsl{Total-internal-reflection within
atmosphere as possible trapping of ray leading to black-hole (for $n
= 3$, $r < 30$)}}}}
\end{figure}
%%%%%%%%%%%%%%%%%%%%%%%%%%%%%%%%%%%%%%%%%%%%%%%%%%%%%%%%%%%%%%%%%%%%%%
For max possible deviation situation when total-internal-reflection
may occur at point $C$ as shown in Figure-$3$; the ray after traveling
a few rounds along the polygon(s) can ultimately come-out from any
vertices including $C$. It may be noted that refraction angle $ r =
(90 - \theta)$ and that $ 2\theta n = 360 $; thus $r = 90 -
\frac{180}{n}$ where $n$ is the number of sides of the polygons shown
in Figure-$3$. The `minimum' possible n for a polygon to exist is $3$
(triangle); so for $n =3$, $r = 30$ degrees. The limiting case for $n =
3$, $r = 30^{0}$ (as shown in Figure-$3.c$) may be looked at as if
this limiting case correspond to black-hole formation. This is
because if the critical angle $ r < 30^{0} $ (i.e., $\mu > 2$), the
light ray will directly fall (be trapped) onto the star-core and
will be absorbed, thus the ray will not come-out. So, black-hole
condition is $n = 3$, $r \leq 30^{0}$ or $\mu \geq 2$ as given in Eqs.
\ref{eq8}, \ref{eq9} \& \ref{eq10}.
\begin{equation}
\label{eq8}
\mu = (\epsilon_{r})^{\frac{1}{2}} = (1 +
\chi)^{\frac{1}{2}} \geq 2
\end{equation}
\begin{equation}
\label{eq9}
\chi \geq 3
\end{equation}
From Eq. \ref{eq4} (with k=2) \& Eq. \ref{eq9}, the final condition
for black-hole is thus given by,
\begin{equation}
\label{eq10}
\frac{GM}{c^{2}R} \geq \frac{3}{4}
\end{equation}
This seems to be in reasonable (middle) agreement with the known
condition $\frac{GM}{c^{2}R} = 1$ (from Newtonian red-shift
approach) or $\frac{GM}{c^{2}R} = \frac{1}{2}$ (from
general-relativity/Schwarzchild-radius), for black-hole
\cite{ref3,ref4}. The present approach also indicates that the
black-hole will have a thick skirt of atmosphere with minimum $\mu =
2$ and $h = R$. Figure-$3$ shows: total-internal-reflection within
atmosphere- as possible trapping of ray and subsequent absorption
into the core, leading to formation of black-hole (for $n=3$; $r <
30^{0}$, $\mu > 2$).\\

The black-hole introduced here through is of new type/class (optical).
It appears that the core of the black-hole is a shrunk heavy neutron-star
surrounded by thick glassy skirt of atmosphere of heavy elements
(including glass-forming Si) from the remnant of the supernovae-explosion.
The trapped radiation (information) can come-out as re-radiation, however differently,
from the black-hole core, agreeing with Hawking \cite{ref30,ref31}
that `black-holes are not so black'. For neutron-star $n > 3$ \&
$\mu < 2$ whereas for black-hole $n = 3$ \& $\mu > 2$.

%%%%%%%%%%%%%%%%%%%%%%%%%%% SECTION 5 %%%%%%%%%%%%%%%%%%%%%%%%%%%%%%%%%%%%
%%%%%%%%%%%%%%%%%%%%%%%%%%%%%%%%%%%%%%%%%%%%%%%%%%%%%%%%%%%%%%%%%%%%%%%%%%%
\section{RED/BLUE SHIFT: NEW ALTERNATIVE EXPLANATION DUE TO REFRACTION-PHENOMENON}

According to Newtonian or Einstein's theory the gravitational red
(or blue) shift $ \frac{d\lambda}{\lambda} = \frac {GM}{c^{2}R} $
and the simple known explanation is said to be as that: when ray
from space approaches towards a massive body the star or planet it
gains gravitational energy, thus frequency $\nu$ increases (blue
shift); but as velocity of light c assumed to be constant, $\lambda$
decreases accordingly.\\

But the physics of the present explanation is quite different. It is
considered/opined (as explained in section $1$, paragraph $2$) that gravitation
does not influence photon; so the photon's energy $h\nu$ hence $\nu$ remains
constant, as is known to be so during refraction. But when from
space (vacuum) the light-ray enters the atmosphere (medium) of star
or planet, the velocity decreases from $c_{o}$ to $c_{m}$, $\nu$
remaining constant, hence $\lambda$ decreases (blue shift). So, blue
shift is explained but the reason is quite different. Similarly,
when ray goes out of atmosphere (medium) to space (vacuum) red-shift
occurs. It may be emphasized that with the present
explanation/theory, the red (or blue) shift is in-fact not
`gravitational red/blue shift' but `refraction red/blue
shift'.\\

The blue/red shift can also be estimated as follows:
$\frac {d\lambda}{\lambda} = \frac{(\lambda_{0} -
\lambda_{m})}{\lambda} = \frac{(\nu\lambda_{o} -
\nu\lambda_{m})}{\nu\lambda_{o}} = 1 - \frac{c_{m}}{c_{o}} = 1 -
\frac{1}{\mu} \sim \frac{(\mu - 1)}{\mu}$. Since for atmosphere $\mu
\sim 1$ and that $ \mu = (\epsilon_{r})^{\frac{1}{2}} = (1 +
\chi)^{\frac{1}{2}} \sim 1 + \frac{\chi}{2}$, the red/blue shift is
given by Eqs. \ref{eq11} \& \ref{eq12} and by Eq. \ref{eq13} (using
(Eqs. \ref{eq12} \& \ref{eq4});
\begin{equation}
\label{eq11}
\frac{d\lambda}{\lambda} = (\mu - 1)
\end{equation}
\begin{equation}
\label{eq12}
= \frac{\chi}{2}
\end{equation}
\begin{equation}
\label{eq13}
= \frac{kGM}{c^{2}R}
\end{equation}
The red/blue shift $\frac{d\lambda}{\lambda}$ predicted by
refraction is $\frac{GM}{c^{2}R}$ from Eq. \ref{eq13} with
fuzz-factor $k = 1$. This shift is in agreement with the known
gravitational shift. It is not inappropriate to use some fuzz-factor
as $k$, in view of inaccuracies/uncertainty in the model/parameters.
The important thing is the `physics' behind the shift; the authors
wish to mention following points:\\

(i) When light enters from space (vacuum) to atmosphere (medium), there
is a blue-shift; and when light goes away from atmosphere to space, there
is a red-shift. This is well in agreement value-wise with Newtonian/Einstein's
gravitational red/blue shift; but authors would like to emphasize
that physics of all the three theories (Newtonian, Einstein's and
the authors' refraction-based present theory) are different.\\

(ii) The present refraction theory of red/blue shift predicts that: once
the ray goes out of atmosphere and travels farther from star, there
is no more further red shift as expected for conventional
gravitational red shift; the red shift only occurs when the light
ray comes out of atmosphere, due to refraction phenomenon. Within
the atmosphere also, there would be some red/blue shift due to
variation in density or refractive-index of the medium.\\

(iii) When, say, for example, light enters from vacuum to atmosphere, both
present and previous theories predict same result blue-shift but the
causes (physics) are different. Also, there is some difference in
present \& previous theories for what is constant \& what varies.
Previous (Newtonian/Einstein's) theories consider velocity of light
$c$ constant, $\nu$ increases (due to gravitational-energy/time-dilation)
thus blue-shift, $\lambda$ decreases to keep c constant. Whereas, present
refraction-based theory considers that energy hence $\nu$ remains constant, $c$
decreases (from $c_{o}$ to $c_{m}$), $\lambda$ decreases (from
$\lambda_{o}$ to $\lambda_{m}$) thus blue-shift. For both (present
\& previous) theories $\lambda$ decreases for blue-shift; but in
previous theories $\nu$ increases \& $c$ remains constant whereas in
present theory $\nu$ remains constant and $c$ decreases as the ray
enters into atmosphere.

%%%%%%%%%%%%%%%%%%%%%%%%%%% SECTION 6 %%%%%%%%%%%%%%%%%%%%%%%%%%%%%%%%%%%%
%%%%%%%%%%%%%%%%%%%%%%%%%%%%%%%%%%%%%%%%%%%%%%%%%%%%%%%%%%%%%%%%%%%%%%%%%%
\section{DISCUSSION AND PREDICTIONS}

The proposed refraction-based explanation quite successfully
explains: (i) bending of light (ii) red/blue shift and (iii) other
aspects such as lensing and black-hole. The authors suggest the
following and there could be possibilities of testing the
novelties.\\

(1) The semi-empirical estimate of equivalent
atmospheric-height (h) near a `star' is roughly $ h =
\frac{GM}{c^{2}}$. But this formula is no good for
atmospheric-height $(h^{/})$ of `planet or satellite', where
radiation pressure is almost absent and important factors are
gravity, pressure \& density. For `planet/satellite' - the formula,
if any, may be entirely different from the formula for `star'.
However, looking for a similar formula and noting that velocity of
sound $c_{s}$ is related to pressure \& density; the equivalent
atmospheric-height for `planet/satellite is empirically suggested/
modified roughly as $h^{/} = \frac{GM}{c c_{s}}$ which gives
reasonable possible values (order of magnitude wise) $ h^{/} =
4~K_m $ for earth and $ h^{/} = 20~m $ for moon; but as we
know, $ h^{/} \sim 0 $ for moon for different reason (escape velocity).\\

(2) As a daring step the authors conclude that gravitational-attraction
is between material bodies only. Thus gravity does not influence/attract
rest-mass-less photon, or photon (electro-magnetic wave) is unaffected
by gravity. Since matter-less photon doesn't has grain-mass \cite{ref32},
it will not have any gravitational or inertial mass either; else it would
have been possible to accelerate the photon  and that all the photons in
a Light-beam could focus by themselves but such an auto-focusing effect
has never been observed \cite{ref5}. \\

(3) Bending of light(photon)-path is
neither due to Newtonian `gravitational-attraction' nor due to
Einstein's `geodesic', but due to refraction-phenomenon of optics
within the atmosphere.\\

(4) There should be no-refraction and
thus no-bending of light around a planet or satellite with almost
no-atmosphere (such as on moon).\\

(5) Black-hole has a thick skirt of atmosphere $(h = R)$ of high
refractive index $(\mu = 2)$. Black-hole physics is - first trapping
of light (due to total-internal-reflection) within atmosphere and
finally absorbed within the black-hole core which can re-radiate
it out in due course.\\

(6) Gravitational-lensing being the true refraction-lensing; should
show some chromatic effect/aberration, which may however be too less
to be noticed normally.\\

(7) The red-shift for example, occurs only when the ray comes out of
atmosphere and no further red-shift afterwards. Frequency $\nu$
remains same, wavelength $\lambda$ and velocity of light $c$ changes.
Some shift within atmosphere also possible, due to possible
variation of $\mu$ within it.\\

(8) The new refraction-based explanation is so obvious that it spares
little room for doubts. In future if the potentials of this new approach is
recognized/appreciated, it would possibly have important bearings
on newer understanding of cosmology from different-perspective.

%%%%%%%%%%%%%%%%%%%%%%%%%%%%%%%  SECTION 7  %%%%%%%%%%%%%%%%%%%%%%%%%
%%%%%%%%%%%%%%%%%%%%%%%%%%%%%%%%%%%%%%%%%%%%%%%%%%%%%%%%%%%%%%%%%%%%%
\section{EINSTEIN'S AND AUTHORS' VIEWS ON CURVATURE OF SPACE-TIME}

In fact due to gravity \cite{ref7}, density thus refractive-index of the
atmosphere varies in radial direction \cite{ref33}; thus during bending, the
light ray actually follows a curved path due to variation of $\mu$
within the medium. This curved path (as shown by a thin free-hand
drawn curved-line in Figure(1.b)) is apparently considered as
Einstein's `geodesic' of general-relativity [1-3] whereas in fact it is
`geodesic' due to refraction through the medium of varying
$\mu$.\\

As per Einstein's general-relativity [1-3] the 4-D empty
space-time is curved (warped) around a mass. The present (authors')
refraction- theory indicates that 3-D space-atmosphere may be
considered curved in view of density-variation (warping) of
atmosphere \cite{ref33} around the mass. With passage of `time' as the light ray
proceeds forward it follows a curved path in the 3-D `space'-
atmosphere creating an impression of `geodesic' in 4-D
space-time.\\

Briefly summarized - (1) Einstein's view and (2)
authors' view on space-time and gravity are as follows:\\

(1) As per Einstein's general-relativity the very `vacuum' of 4-D
space-time is warped/curved around a mass; also that there is no
gravity but apparently appears due to curvature of the space-time.
`Space-time curvature (warping) causes apparent-gravity'.\\

(2) As per authors' view the gaseous-`atmosphere' in the 4-D space-time is
warped/curved around the mass; and that gravity is very much there
and the warping/curving (variation) of atmospheric properties is infact due
to gravity. `Gravity causes density variation (warping/curving) of
space-time'. The space-time `fabric' which warps (curves) around the
mass is not `vacuum' but the `atmosphere' \cite{ref33}. Warping of
the fabric of space-time, is in fact, is the variation of properties
(density and refractive-index) within the surrounding-atmosphere
around the massive body.

%%%%%%%%%%%%%%%%%%%%%%%%%%%%%%%%%%%%%%%%%%%%%%%%%%%%%%%%%%%%%%%%%%%%%%%%%%
%%%%%%%%%%%%%%%%%%%%%%%%%%% SECTION 8 %%%%%%%%%%%%%%%%%%%%%%%%%%%%%%%%%%%%
\section{`CURVATURE CAUSES GRAVITY' OR `GRAVITY CAUSES CURVATURE'?}

A very important question, like egg \& chicken famous-problem, can be asked
that whether (i) is it the space-time curvature around big-mass that
causes apparent gravity-effect or (ii) is it the real gravity-force
that causes atmospheric density variation (viz. curvature) around
the big-mass ? As explained by Gupta in a satiric paper \cite{ref33} it is concluded
that: as per Einstein's general-relativity [1-3] it is the first
answer (Curvature causes Gravity-effect), whereas as per Gupta's
gravity-theory \cite{ref7} it is the second answer (Gravity causes
Curvature-effect). In fact \cite{ref33} gravity force is real and
not the artifact of curvature, rather curvature (density-variation)
can be said as an effect of gravity. Origin of gravity is not the
curvature, but it is the gravity which is the origin of the
so-called curvature. Einstein's space-time curvature of empty-space
is somewhat equivalent to density (or refractive-index) variation of the
atmosphere in matter-universe; that in a way implies equivalence
\cite{ref33} of bending of light near a star (i) due to curvature of
warped empty-space  implying general-relativity based
gravitational-bending or (ii) due to variation of density
(refractive-index) causing refraction which makes the light-ray to
bend \& follow a curved path (as shown by the thin curved-line
within the atmospheric-region in Figure $(1.b)$). The two differently
appearing explanations seem to have something in common, probably
these two are almost (more-or-less) the same thing appearing
different as being viewed from different perspective.

%%%%%%%%%%%%%%%%%%%%%%%%%%%%%%%%%%%%%%%%%%%%%%%%%%%%%%%%%%%%%%%%%%%%%%%%%%
%%%%%%%%%%%%%%%%%%%%%%%%%%% SECTION 9 %%%%%%%%%%%%%%%%%%%%%%%%%%%%%%%%%%%%
\section{SOME TESTS FOR THE REFRACTION-BASED THEORY}

Early sun-rise and late sun-set  than the actual-happenings (say, by
about 3 minutes) are the well known \& established facts
\cite{ref34}, which are known to occur due to refraction-based
bending of light in the earth's atmosphere. But  general-relativity
based gravitational-bending of star-light near Sun is considered to
be different, based on phenomenon of space-time curvature. Though
the proposed refraction-based theory in the present-paper seems to
has more in-common than generally realized, but the `physics' is
different. Some possible litmus-tests are proposed, as follows, to
test/verify the herein proposed refraction-based theory for
terrestrial \& cosmological phenomena. Not much details, however,
for the experimental schemes for the proposed tests have been
provided; but the ideas \& the salient features behind the tests
have been clearly indicated.\\

(1) General-relativity based gravitational-bending of light is
frequency-independent whereas the refraction-based bending is quite
obviously frequency-dependent. For example: it is well known that in
the VIBGYOR-spectrum, the higher-frequency blue-light bends more than
the lower-frequency red-light. The frequency- dependent bending of light
near a star, though may be small, if really found in terrestrial/cosmological
observations, will establish the truth of refraction-based theory.
Also, it is expected that there would be more bending of the
high-frequency X-ray than the low-frequency radio-signals existing
within the wide-spectrum of the star-light.\\

(2) If the gravitational-lensing is a result of  refraction-based theory which
is frequency-dependent, the sky  therefore  should be more colorful
in view of chromatic - aberration in the images, which may possibly
be noticed/observed.\\

(3) As explained in the section-5 of this paper, it is explained that the
red/blue shift is in fact due to the fact that light-speed in denser-medium
is less than that in rarer-medium. It is well known that the frequency of
the incident light remains the same during refraction, hence any blue/red shift
is in-fact due to corresponding change of wavelength $(\lambda =
\frac{c}{\nu})$. With some specially designed experiment, it could
possibly be found that, say, blue-shift is `not due to real-increase
in frequency but due to real-decrease of wavelength' of light in the
medium wherein light-speed is lesser.\\

(4) If we fabricate a sphere of radius $2R$ in such a way, that it
has a dark-black light-absorbing inner-core of radius $R$ surrounded by
a glass-like material of $\mu \geq 2$ of thickness R all around; the
whole sphere would appear black (light would be entrapped in it, as
explained in section-$4$). This would make a sort of black-hole type
object which could be held in hand (of course without any danger!).
Diamond has $\mu > 2$, but it seems costly \& impracticable to use
diamond for it. Is there any other such glass-like material $(\mu \geq 2)$
which could be affordable \& practicable to be used to construct such toy
type mini black-hole? \\

(5) In the empty (near vacuum) space little-away from a star, galaxy or
black-hole, there should be no refraction hence no bending of light. Indeed,
such fact, is reported by Dowdye \cite{ref35,ref36} that:`no light bending
occurs in empty space just above the plasma rim of Sun'.
%%%%%%%%%%%%%%%%%%%%%%%%%%%%%%%%%%%%%%%%%%%%%%%%%%%%%%%%%%%%%%%%%%%%%%%%%%
%%%%%%%%%%%%%%%%%%%%%%%%%%% SECTION 10 %%%%%%%%%%%%%%%%%%%%%%%%%%%%%%%%%%%%
\section{CONCLUSIONS}

It is suggested that - Gravitation is only between material bodies
and that the zero-rest-mass photon is unaffected by gravity. The
alternative novel approach to explain phenomena such as bending of
light near a star and gravitational red/blue shift is based on
refraction phenomenon of optics. Bending of light is due to bending
of ray due to refraction within the star's atmosphere. The red/blue
shift is due to optical-phenomenon of change of wavelength
(frequency remaining same) due to change in velocity of light in the
atmospheric medium. Other aspects such as black-hole and
gravitational-lensing are also re-examined in the new perspective of
refraction-phenomenon. Interesting predictions are made and also a
few tests are suggested. In fact many of the
general-relativity-tests are explained {\it without} general-relativity on
the basis of refraction-phenomenon. The new approach could have
important bearing on newer understanding of space-time, gravity and
cosmology from different perspective.

%%%%%%%%%%%%%%%%%%%%%%%%%%%%%%%%%%%%%%%%%%%%%%%%%%%%%%%%%%%%%%%%%%%%%%%%%%%%%%%%%%%%%%%%%%%%%%%%%
%%%%%%%%%%%%%%%%%%%%%%%%%%%%%%%%%%%%%%%%%%%%%%%%%%%%%%%%%%%%%%%%%%%%%%%%%%%%%%%%%%%%%%%%%%%%%%%
\section*{{\it Acknowledgements}}
The authors are thankful to Dr. V. B. Johri, Dr. B. Das and Dr. M. S. Kalara for advice \& help.
I.E.T., AKTU(UPTU), Lucknow, IUCAA, Pune, I.I.T., Kanpur and GLA University, Mathura are also
thanked for direct/indirect support.
%%%%%%%%%%%%%%%%%%%%%%%%%%%%%%%%%%%%%%%%%%%%%%%%%%%%%%%%%%%%%%%%%%%%%%%%%%%%%%%%%%%%%%%%%%%%%%%%%%%%%%%%%
%\newline
%\nonumsection{References}

\end{document}